# Spatiotemporal interaction of optical beams in bi-dispersive media

Nikolaos Moshonas[1,*], Yannis Kominis[1], Panagiotis Papagiannis[1], Kyriakos Hizanidis[1] and Demetrios N. Christodoulides[2]

[1]*School of Electrical and Computer Engineering, National Technical University of Athens, 157 73 Athens, Greece*

[2]*College of Optics/CREOL, University of Central Florida, Orlando, Florida 32816-2700, USA*

[*]*Corresponding author: nmoshon@central.ntua.gr*

## Abstract

The interaction of optical beams in bi-dispersive nonlinear media with self-focusing nonlinearity is numerically investigated. Under proper conditions, the interaction of spatially separated beams can lead to the creation of two prominent filaments along the other spatial or temporal dimension and, thus to an effective beam exchange. This X-wave generation-based effect could potentially be exploited in optical realizations of spatial or spatio-temporal filtering.

The propagation of narrow optical beams in media with self-focusing Kerr nonlinearity has been the focus of intense investigation for many years [1, 2]. In the case where two-dimensional diffraction and anomalous dispersion are present, self focusing (SF) leading to full wave collapse can be observed, if the initial power of the beam exceeds a critical value $N_c$ [3]. More recently interest arose in Kerr nonlinear media with normal dispersion where the SF process can be delayed and collapse can be arrested [4-7]. Instead of collapse, pulse splitting may occur along the temporal dimension. This effect is associated with the phenomenon of conical emission. The latter effect is related to the generation and amplification of spectral sidebands that obey a dispersion relation that demands $k_\perp^2 \propto \Omega^2$, which represent hyperbolas bound by the cone $|k_\perp| = \pm\sqrt{kk''}\Omega$ (with $k$ being the wavenumber of the beam) [6, 8]. This relationship spreads the energy along these "spectral hyperbolas" and strengthens the dispersion factor which, in turn, redirects the energy away from the center of the wavepacket, preventing the collapse and forming two light filaments, (one blue shifted and the other red shifted), thus, splitting the beam. We note that these "spectral hyperbolas" are the same as those involved in spatiotemporal modulational instability (MI), as far as the spectrum of a perturbed continuous wave (CW) is concerned [9]. MI is also responsible for the spontaneous generation of nonlinear X-waves [10]. These spontaneously generated X-shaped light bullets have been observed in lithium triborate χ(2) crystals [11]. In another recent work, X-wave generation was considered by exploiting the interaction of a 2D localized Gaussian wavepacket and a CW background [12, 13]. It was found that the X-wave generation process depends on both the relative phase and amplitude of the background with respect to the superimposed wavepacket [14, 15]. In fact it was the MI of the CW which enabled the creation of these X-like structures.

In the present investigation we consider the interaction of two 2D localized wavepackets, spatially separated, on top of a broad CW beam. Even though X-wave generation is not the aim of this work, the MI generated X-like structures are omnipresent. It is thus observed, that their interaction can lead to the emergence of two dominant light filaments with temporal separation. In the more general setting of a bi-dispersive nonlinear medium [13], the temporal dimension is replaced by the other spatial dimension and the aforementioned filaments are now separated spatially, though transversely to initial (at launching) spatial separation of the beams. In Fig. 1 surfaces of constant light intensity in two settings of practical interest are presented, along with a conceptual realization: In Fig.1a, two light beams spatially separated at the launching point ($z=0$) end up temporally separated at the end ($z=5$). In Fig.1b is merely demonstrated that SF does not allow space-time exchange, i.e., initial temporal separation does not lead to spatial separation. In Fig. 1c a possible experimental arrangement is sketched: Two light beams which are simultaneously incident at an angle at two spatially separated ports generate a temporal sequence of two light pulses emanating from the center at the end of the slab waveguide; a weak continuous wave (CW) is also injected at the launching point.

In the aforementioned configuration (planar normally dispersive waveguide), beam propagation can be described by the following nonlinear Schrödinger equation (NLSE),

$$i\frac{\partial u}{\partial z} + \frac{\partial^2 u}{\partial x^2} - \frac{\partial^2 u}{\partial t^2} + |u|^2 u = 0 \qquad (1)$$

where, $z = Z/z_0$; $t = (T - Zk')/t_0$ and $x = X/w_0$ are the normalized longitudinal, time and transverse coordinates, $u$ is the normalized electric field given by $u = E/I_0^{1/2}$, with $E$ being the envelope amplitude and $I_0$ the characteristic intensity. As a plausible example, we consider an AlGaAs waveguide operated at $\lambda=1.55\mu m$ with a linear refractive index $n=3.34$. The nonlinear coefficient of this system is $n_2 = 1.5 \times 10^{-13} cm^2/W$ and the normal dispersion parameter is $k'' = 1.35 \times 10^{-24} s^2/m$. In the examples to follow the beams have a spatial width $w_0$=10.5μ$m$. Hence $z_0$=*3mm*, $t_0$=*45fs* and $I_0$=*5.5W/μm²*. On the other hand, the intensity used for the CW is between 1-4% of the maximum beam intensity. The effective horizontal length (*x* dimension) of the waveguide is not of importance since it is considered very long, its transverse (*y* dimension) width is taken here to be 1μ$m$ and its length is *Z=15mm* (*z=5*). The relationship of the diffraction and dispersion effects was taken under consideration through the relationship of their respective lengths $L_{df} = w_0^2 k/2$, $L_{ds} = t_0^2/k''$. Their aspect ratio is of major importance, and is kept $L_{df}/L_{ds} = 1/2$ throughout this work.

The evolution of *u* is greatly affected by its "mass", defined as $N = \int |u|^2 \, dxdt$. In the case of a single beam propagating in an anomalous dispersive medium, with a power that exceeds a certain critical value $N_c$=*4π*, the end result is self focusing and collapse. Bergé et al. [15], in their key work concerning beam interaction, consider the respective initial beam power $N_i$, in combination with their spatial separation, as criterion for merging and probable focusing. Here we follow the same guidelines, taking into consideration the role of CW and the features of normal dispersion. On the other hand, SF, which is followed by beam splitting, occurs in media with normal dispersion, for powers that are usually greater than $N_c$, and is related to the aspect ratio of the diffraction and dispersion lengths [5]. Since here we assume the coexistence of two localized Gaussian beams along with a CW background, we can consider *u* as a superposition of the beams $u_1$ and $u_2$ with $u_{CW}$, namely, $u = u_1 + u_2 + u_{CW}$. It is assumed that the "mass" variation, $\Delta N_i$, generated by the interaction of the beams with the CW is incorporated in the individual beam themselves (itself). This assumption can be easily met if one considers the relatively lower power level of the CW or quasi-CW beam. Hence, one may set, $u=u_1+u_2+u_{CW}$ where,

$$u_{1,2} = A\exp\left[-\frac{(x\mp x_0)^2+(t-t_0)^2}{2} \pm i\Delta k_x x + i\phi_{1,2}\right], \qquad u_{CW} = Aa_{CW}\exp(-i\varphi) \qquad (2)$$

where $A$ is the amplitude of the beams, $\Delta x = 2x_0$, $\Delta t = 2t_0$ are the initial spatial and temporal wavepacket separations, $\Delta k_x$ is the initial transverse wavenumber mismatch, $\phi_1$ and $\phi_2$ are the initial phases of the beams, $\alpha_{CW}$ is the relative amplitude of the CW and $\varphi$ its phase. Spatiotemporal integration of the individual modified "masses", $|u_{1,2}+u_{CW}|^2$, approximately yields $N_i + \Delta N_i = N_i\left[1+4a_{CW}\cos(\Delta\phi_i + \Delta k_x \Delta x/2)\exp\left(-|\Delta k_x|^2/2\right)\right]$, where, $\Delta\phi_i = \phi_i - \varphi$, ($i$=1,2) is the phase difference of the respective beam with the CW. This, in turn, should change the individual beam power necessary for SF and splitting and can affect also the process of merging and the probable SF thereafter. For the value of the aspect ratio between the diffraction and dispersion lengths, considered here, the necessary power that can trigger individual SF of a wavepacket is not $N_c$. Actually it is more than $2N_c$ [5]. The interaction of the beams depends on their relative phase but also on the individual powers [15, 16]. The numerical investigation that follows takes in to consideration all of the above. In Figs. 2-4 the horizontal axis corresponds to the spatial ($x$) dimension and has a length of 50 transverse pulse widths (-263$\mu m$ to 263$\mu m$) and the vertical axis corresponds to the temporal ($t$) dimension at a length of 50 pulse widths (-1.12$ps$ to 1.12$ps$).

The NLSE is solved via a beam propagation method. In all cases the spatiotemporal window is taken broad enough to avoid artificial reflections from the boundaries. The CW is retained throughout this investigation since its presence retards the diffraction of the beams and helps to the concentration of the energy in relation with the MI created structures [12]. Furthermore, without any loss of generality we also set $\varphi_1=\varphi_2$. In Fig.2, the output intensity profiles are shown at $z=5$ (at 15$mm$) when two beams are lunched simultaneously, with $\Delta x = 4$ (42$\mu m$ for the reference example), $\Delta\phi_i = 0$, $\Delta k_x = 0$, and in the presence of a CW with $a_{CW} = 0.1$, are shown. The initial beam powers are varying, and in Fig. 2a $N_i$=0.9$N_c$, which is lower than the minimum needed power for the beams to merge and focus [16]. That should not happen, no matter how close the two beams might start. In this case, the beams diffract, but the energy is weakly concentrated in a low central lobe and along hyperbolic curves, resembling an X-like structure. For beams with $N_i$=1.85$N_c$, (Fig.2b) their power is high enough that along with the part offered by the CW, force them to start individual splitting. Nevertheless, soon after this splitting starts, the filaments merge along the temporal axis. However, the energy is not concentrated in two prominent lobes but in many smaller ones. For even higher beam power ($N_i$=3$N_c$) the result is multiple splitting and filamentation (Fig.2c).

Next we investigate the effect of initial spatial separation of the beams. Beam power is set at $N_i$=1.5$N_c$ and the CW is of higher amplitude than before, namely $a_{CW} = 0.2$. Thus, the power for individual SF and splitting is actually close to 1.5$N_c$. For $\Delta x = 3$, in Fig.3a, the two beams tend to temporally split, but they soon merge. Two individual filaments of opposite frequency sidebands emerge and draw away from each other, but many other smaller peaks appear. For $\Delta x = 4$, the beams split individually, but the filaments merge and concentrate at the t-intercept points of the first hyperbola. This is a more favorable result, since these filaments have significant more power than the rest that are being created, but they emerge rather late, around $z$=4. They move away along t-axis and retain their amplitude until $z$=5. For $\Delta x = 5$ we observe individual splitting. Diffraction spreads the energy, mainly along the x-centered hyperbolas.

We next consider how the spatial wavenumber mismatch between the beams and their phase difference with the CW, affect the beam evolution. The relative phase between the beams is set to zero, $a_{CW}=0.2$, $\Delta x=4$ and $N_i=2N_c$. Figure 4 presents the output profiles for $\Delta k_x=0$ [Figs.4(a-c)] and *0.2* [Figs.4(d-f)] and *φ=0* [Figs.4(a,d)], *π/2* [Figs.4(b,e)] and *π* [Figs. 4(c,f)]. For *φ*=0, the CW adds a lot of energy to the beams resulting in individual splitting along time and generation of many small lobes of nearly equal amplitude. The higher the wavenumber mismatch, the more "momentum" the initial beams have and the more obvious the merging and multiple filamentation are [Figs.4(a,d)]. For *φ*=π/2, the CW does not add any amount of energy to the beams, at least initially. The beams do not self- focus, initial power is not enough for this to happen. Instead, they disperse, but finally merge at the center and split twice. Four filaments are formed, of relatively equal amplitude, which they retain. The case of *φ*=π gives the most efficient intensity reallocation [Figs.4(c,f)]. In this case the CW takes away power from the initial beams. Also, the less their relative wavenumber mismatch is, the more power is "subtracted" from them. The beams disperse quickly, moving to create two filaments of their own along the temporal axis, but then they finally merge in one central lobe which splits in two other lobes (around *z*=2.5) with their peaks positioned at *x*=0 and moving along time, thus emerging at *z=5* clearly separated in time. Although convergence of the beams ($\Delta k_x>0$) does not affect the outcome strongly, even better results are attained for slightly converging beams (Fig.4f). In Fig.1a this latter more favorable case is shown: An initial spatial separation $\Delta x=4$ (*42μm* for the reference example) lead to a temporal separation $\Delta t=16$ (*0.7ps* for the reference example). Furthermore, in Fig.5 the spatial [Fig.5(a-c)] as well as the temporal [Fig.5(a-c)] evolution are shown for the respective to Figs.4(d-f) cases.

In conclusion, it has been shown that the interaction between two spatially separated localized wavepackets and a CW background can lead to the creation of two temporally distinct filaments. The propagation distance where the filaments emerge, as well as their initial position on the time axis, their intensity and robustness depend on the initial characteristics of the interacting beams, as well on these of the CW. The above observations indicate that a controllable all-optical spatial or spatio-temporal switching and filtering is possible in such systems.

# References

1. J. J. Rasmussen, K. Rypdal, "Blow-up in nonlinear Schrödinger equations-I. A general review", Phys. Scripta **33**, 481-497 (1986).

2. L. Bergé, "Wave collapse in physics: Principles and applications to light and plasma waves", Phys. Reports **303**, 259-370 (1998).

3. Y. Kivshar, D. E. Pelinovsky, "Self-focusing and transverse instabilities of solitary waves", Phys. Reports **331**, 117-194 (2000).

4. P. Chernev, V. Petrov, "Self-focusing of light pulses in the presence of normal group-velocity dispersion", Opt. Lett. **17** (3), 172-174 (1992).

5. G. G. Luther, J. V. Moloney, A. C. Newell, "Self-focusing threshold in normally dispersive media", Opt. Lett. **19** (12), 862-864 (1994).

6. G. G. Luther, J. V. Moloney, A. C. Newell, E. M. Wright, "Short-pulse conical emission and spectral broadening in normally dispersive media", Opt. Lett. **19** (11), 789-791 (1994).

7. J. K. Ranka, R. W. Schirmer, A. L. Gaeta "Observation of pulse splitting in nonlinear dispersive media", Phys. Rev. Lett. **77** (18), 3783-3786 (1996).

8. A. G. Livak, V. A. Mironov, E. M. Sher, "Regime of wave-packet self-action with normal dispersion of the group velocity", Phys. Rev. E **61** (1), 891-893 (2000).

9. L. W. Liou, X. D. Cao, C. J. McKinstrie, G. P. Agrawal, "Spatiotemporal instabilities in dispersive nonlinear media", Phys. Rev. A **46** (7), 4202-4208 (1992).

10. C. Conti "X-wave mediated instability of plane waves in Kerr media", Phys. Rev. E **68**, 016606 (2003).

11. P. Di Trapani, G. Valiulis, A. Piskarkas, O. Jedrkiewicz, J. Trull, C. Conti, S. Trillo, "Spontaneously generated X-shaped light bullets", Phys. Rev. Lett. **91** (9), 093904 (1-4), (2003).

12. Y. Kominis, N. Moshonas, P. Papagiannis, K. Hizanidis, D. N. Christodoulides, "Continuous wave controlled nonlinear X-wave generation", Opt. Lett. **30** (21), 2924-6 (2005).

13. D. N. Christodoulides, N. K. Efremidis, P. Di Trapani, and B. A. Malomed, "Bessel X waves in two- and three-dimensional bidispersive optical systems," Opt. Lett. **29**, 1446-1448 (2004).

14. N. N. Akhmediev, S. Wabnitz, "Phase detecting of solitons by mixing with continuous wave background in an optical fiber", J. Opt. Soc. Am. B **9** (2), 236-242 (1992).

15. Y. Kominis, K. Hizanidis, "Solitary wave interactions with continuous waves", Inter. J. Bifurc. Chaos **16** (6), 1753-1764 (2006).

16. L. Bergé, M. R. Schmidt, J. J. Rasmussen, P. L. Christiansen, K. Ø. Rasmussen, "Amalgamation of interacting light beamlets in Kerr-type media", J. Opt. Soc. Am. B **14** (10), 2550-62 (1997).

17. G. I. Stegeman, M. Segev, "Optical spatial solitons and their interactions: Universality and diversity", Science 286, 1518 (1999).

**Figure Captions**

**Figure 1**  Iso-intensity plots at *80%* of both the beams intensity: (a) Appearance of two prominent filaments on the temporal axis after the interaction of two spatially localized beams (the same case is also presented in Fig.4d); (b) two initially temporally localized beams lead to main filaments on the temporal axis; (c) two light beams simultaneously incident at an angle at two spatially separated ports of a conceptual device lead to a temporal sequence of two light pulses emanating from the single center port at the end of the device.

**Figure 2**  Output intensity profile, at *z=5* (*15mm*), with $a_{CW}$=0.1, $\varphi_1=\varphi_2$, $\varphi=0$ and *Δx=4*. The input power and the spatial separation was initially set at (a) $N_i=0.9N_c$, (b) $N_i=1.85N_c$, (c) $N_i=3N_c$.

**Figure 3**  Output intensity profile, at *z=5* (*15mm*) for initial spatial separations: (a) *Δx=3*, (b) *Δx=4*, (c) Δx=5. In all cases, $a_{CW}$=0.2, $\varphi_1=\varphi_2$, $\varphi=0$ and $N_i=1.5N_c$.

**Figure 4**  Output intensity profile, at *z=5*, in the presence of a CW, with $a_{CW}$ =0.2, for various values of its phase (*φ*) and for various values of the wavepackets initial transverse wavenumber difference ($\Delta k_x$); (a, d) *φ=0*, (b, e) *φ=π/2*, (c, f) *φ=π*; (a, b, c) $\Delta k_x = 0$, (d, e, f) $\Delta k_x = 0.2$. In all cases, initial values were set to $N_i=2N_c$, *Δx=4* and $\varphi_1=\varphi_2$.

**Figure 5**  Spatial (a,b,c) and temporal (d,e,f) intensity profiles during propagation for $\Delta k_x = 0.2$, $a_{CW}$ =0.2 $N_i=2N_c$, $\varphi_1=\varphi_2$, *Δx=4* and: (a,d) *φ=0*, (b,e) *φ=π/2*, (c,f) *φ=π*.

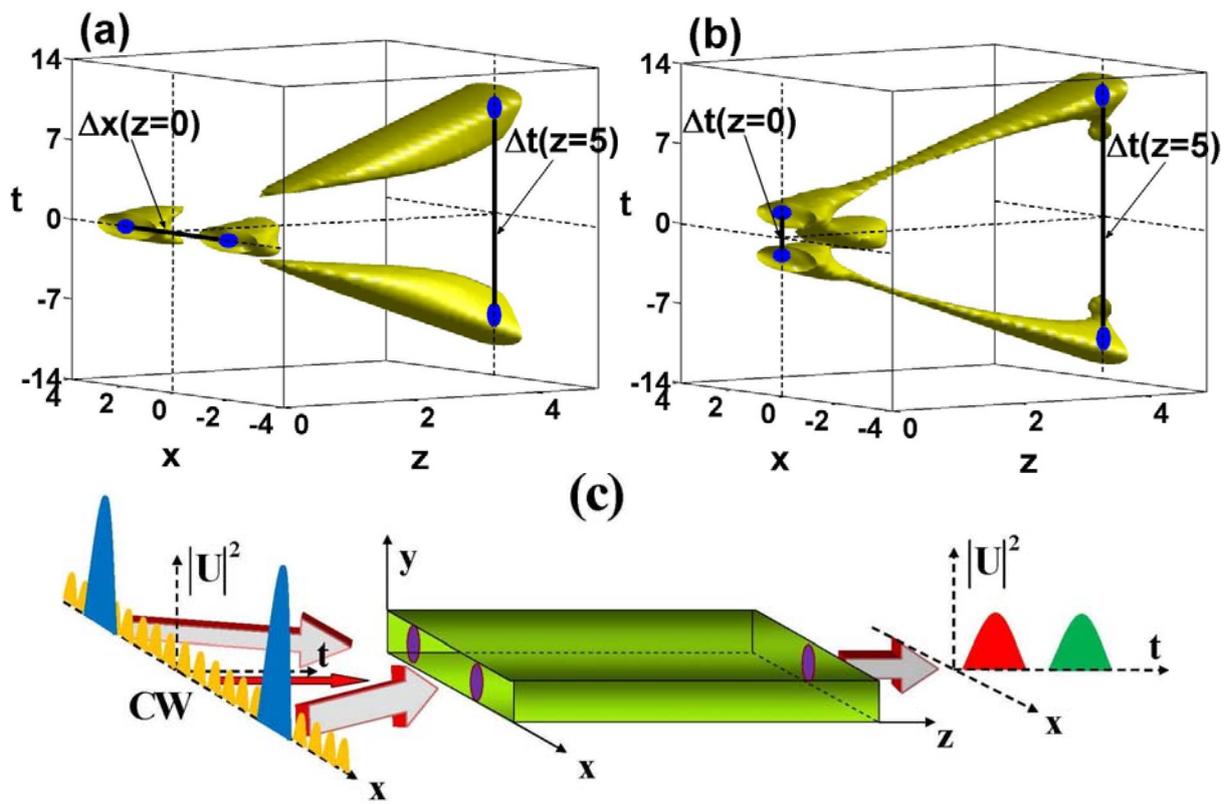

**Fig. 1**

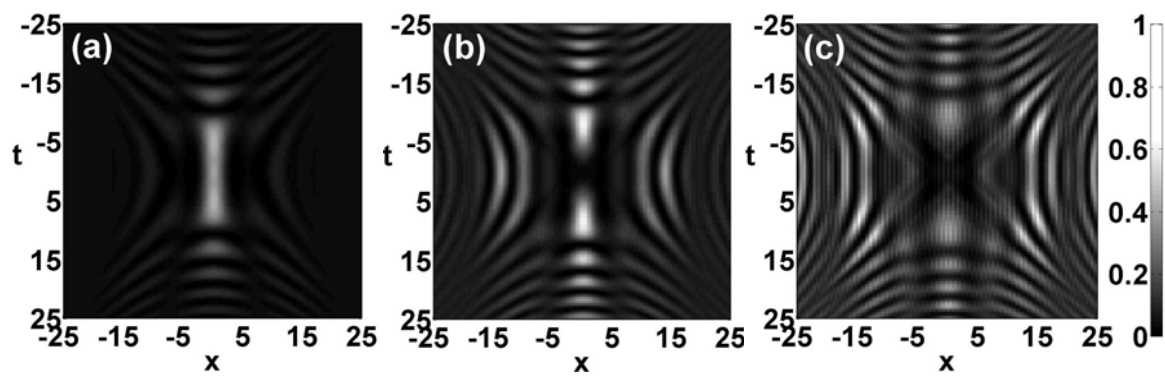

**Fig. 2**

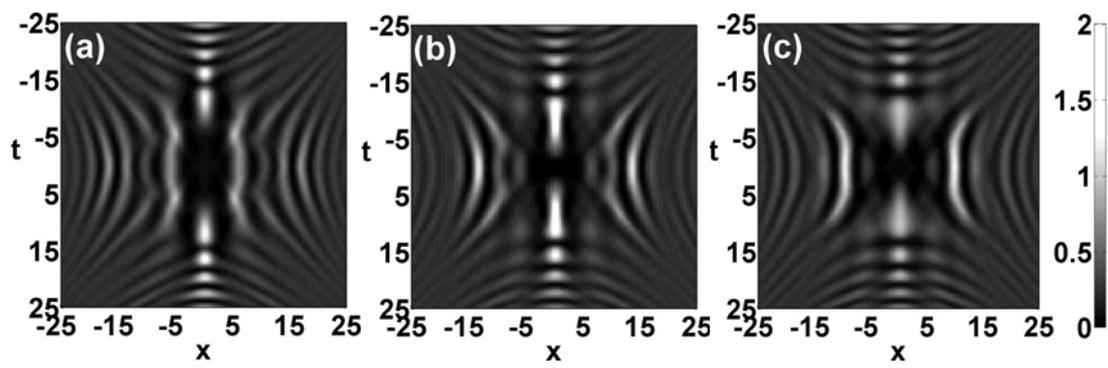

**Fig. 3**

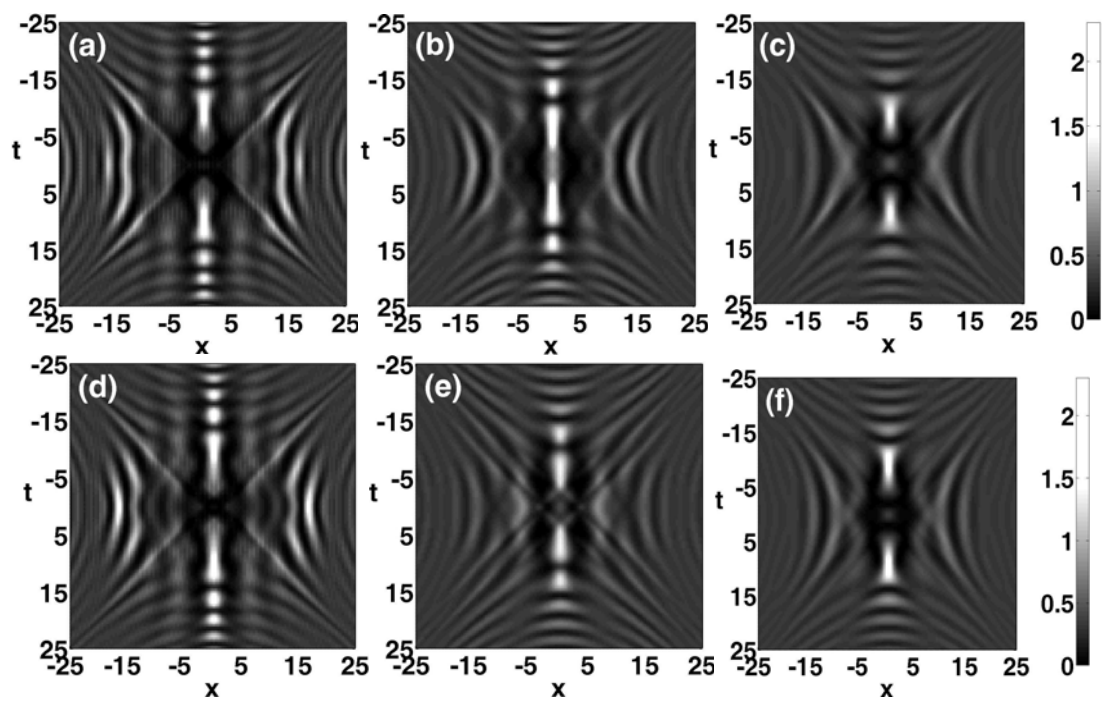

**Fig. 4**

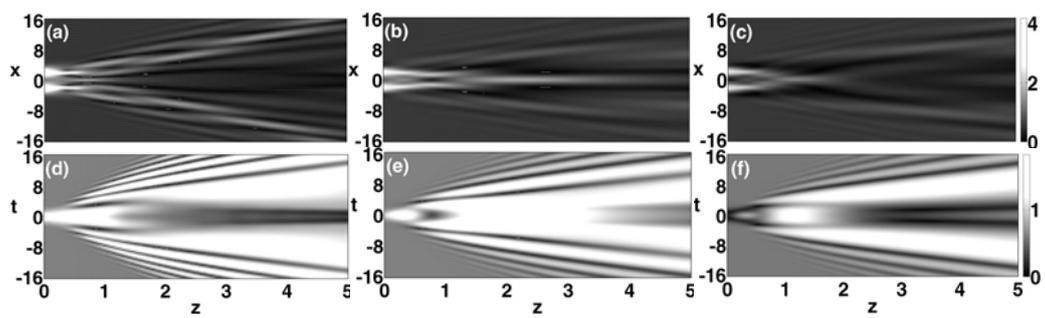

**Fig. 5**